\documentclass[fleqn,10pt]{wlscirep}
\usepackage[utf8]{inputenc}
\usepackage[T1]{fontenc}
\usepackage{setspace} 
\usepackage{xcolor}
\title{Fermi and eRosita bubbles as relics of the past activity of the Galactic black hole}

\author[1,2,3,*]{H.-Y. Karen Yang}
\author[4]{Mateusz Ruszkowski}
\author[5]{Ellen G. Zweibel}
\affil[1]{Institute of Astronomy, National Tsing Hua University, Hsinchu, 300013, Taiwan R.O.C.}
\affil[2]{Department of Physics, National Tsing Hua University, Hsinchu, 300013, Taiwan R.O.C.}
\affil[3]{Physics Division, National Center for Theoretical Sciences, Taipei, 106017, Taiwan R.O.C.}
\affil[4]{Department of Astronomy, University of Michigan, Ann Arbor, 48109, U.S.A.}
\affil[5]{Department of Astronomy, University of Wisconsin-Madison, Madison, 53706, U.S.A.}

\affil[*]{hyang@phys.nthu.edu.tw}

\begin{abstract}

{\bf The newly launched X-ray satellite, eRosita, has recently revealed two gigantic bubbles extending to $\sim 80^\circ$ above and below the Galactic center. The morphology of these ``eRosita bubbles'' bears a remarkable resemblance to the Fermi bubbles previously discovered by the Fermi Gamma-ray Space Telescope and its counterpart, the microwave haze. The physical origin of these striking structures has been intensely debated; however, because of their symmetry about the Galactic center, they likely originate from some energetic outbursts from the Galactic center in the past. Here we propose a theoretical model in which the eRosita bubbles, Fermi bubbles, and the microwave haze could be simultaneously explained by a single event of jet activity from the central supermassive black hole a few million years ago. Using numerical simulations, we show that this model could successfully reproduce the morphology and multi-wavelength spectra of the observed bubbles and haze, which allows us to derive critical constraints on the energetics and timescales of the outburst. This study serves as an important step forward in our understanding of the past Galactic center activity of our Milky Way Galaxy, and may bring valuable insights into the broader picture of supermassive black hole-galaxy co-evolution in the context of galaxy formation.}

\end{abstract}
\begin{document}

\flushbottom
\maketitle
\thispagestyle{empty}

%\noindent Please note: Abbreviations should be introduced at the first mention in the main text – no abbreviations lists. Suggested structure of main text (not enforced) is provided below.

\section*{Introduction}

Recent data taken by the {\it eRosita} satellite has revealed striking images of two giant X-ray bubbles extending $\sim 80^\circ$ (which corresponds to $\sim 15$ kpc assuming a distance to the Galactic Center (GC) of 8.5kpc) above and below the GC \cite{eRosita}. Despite the larger extent, the morphology of the ``eRosita bubbles'' is remarkably similar to the {\it Fermi} bubbles, the two gamma-ray bubbles detected by the {\it Fermi} Gamma-ray Space Telescope in 2010 \cite{Su10}. It was also shown previously that the gamma-ray bubbles have counterparts in the microwave band, known as the microwave/{\it WMAP}/{\it Planck} haze \cite{Finkbeiner04, PlanckHaze}, as well as polarized lobes in radio\cite{Carretti13}. The edges of the {\it Fermi} bubbles at low latitudes also align with earlier detections of X-ray shells in the 1.5 keV band by ROSAT\cite{BlandHawthorn03, BlandHawthorn19}. Because of their morphological similarities, enormous physical sizes, and symmetry about the GC, these fascinating structures most likely originate from the same event of powerful energy outburst from the GC sometime in the past. Understanding their physical origin would therefore provide valuable information about the history of our Milky Way Galaxy.    

Before the detection of the {\it eRosita} bubbles, the physical origin of the {\it Fermi} bubbles and microwave haze had been hotly debated \cite{Yang18}. Key questions to address include whether the gamma-ray emission is hadronic or leptonic, and whether the triggering mechanism is associated with a nuclear starburst, or with activity of the central supermassive black hole (SMBH). Because of the proximity to the GC, the ample spatially-resolved, multi-wavelength observational data provide very stringent constraints on the proposed theoretical models. One strong constraint comes from the hard spectrum of the gamma-ray bubbles with spectral index of $-2.1$ \cite{Ackermann14}, which means that the gamma-ray emission has to be generated by cosmic-ray (CR) protons transported by starburst or active galactic nucleus (AGN) winds that do not easily cool (i.e., the ``hadronic wind models") \cite{Crocker15, Mou14}, by CR electrons quickly transported by AGN jets before they cool (i.e., the ``leptonic jet models") \cite{Guo12a, Yang12, Yang13, Yang17}, or by CRs which are accelerated in situ by shocks or turbulence near the shock fronts (i.e., the ``in situ acceleration models") \cite{Cheng11, Sarkar17, Mertsch19}. Previous studies have found that purely hadronic models are disfavored because the predicted microwave emission from secondary particles produced via pion decay is not enough to account for the observed haze emission \cite{Ackermann14}. In addition, the upper limits in TeV gamma-rays obtained by {\it HAWC} have ruled out hadronic models with a power law extending to PeV energies \cite{HAWC17}. On the other hand, both the leptonic jet models and in situ acceleration models remain promising mechanisms to explain the gamma-ray bubbles and microwave haze \cite{Yang17, Sarkar17, Mertsch19}. In this work, we show that the new {\it eRosita} data provides crucial information that allows us to put additional constraints on these two scenarios, and that the combination of the gamma-ray, X-ray, and microwave images and spectra strongly suggest that past jet activity of the GC black hole is the likely culprit.

\section*{Simulations of Past Jet Activity of Sgr A*}

\begin{figure}[ht]
\centering
\includegraphics[width=\linewidth]{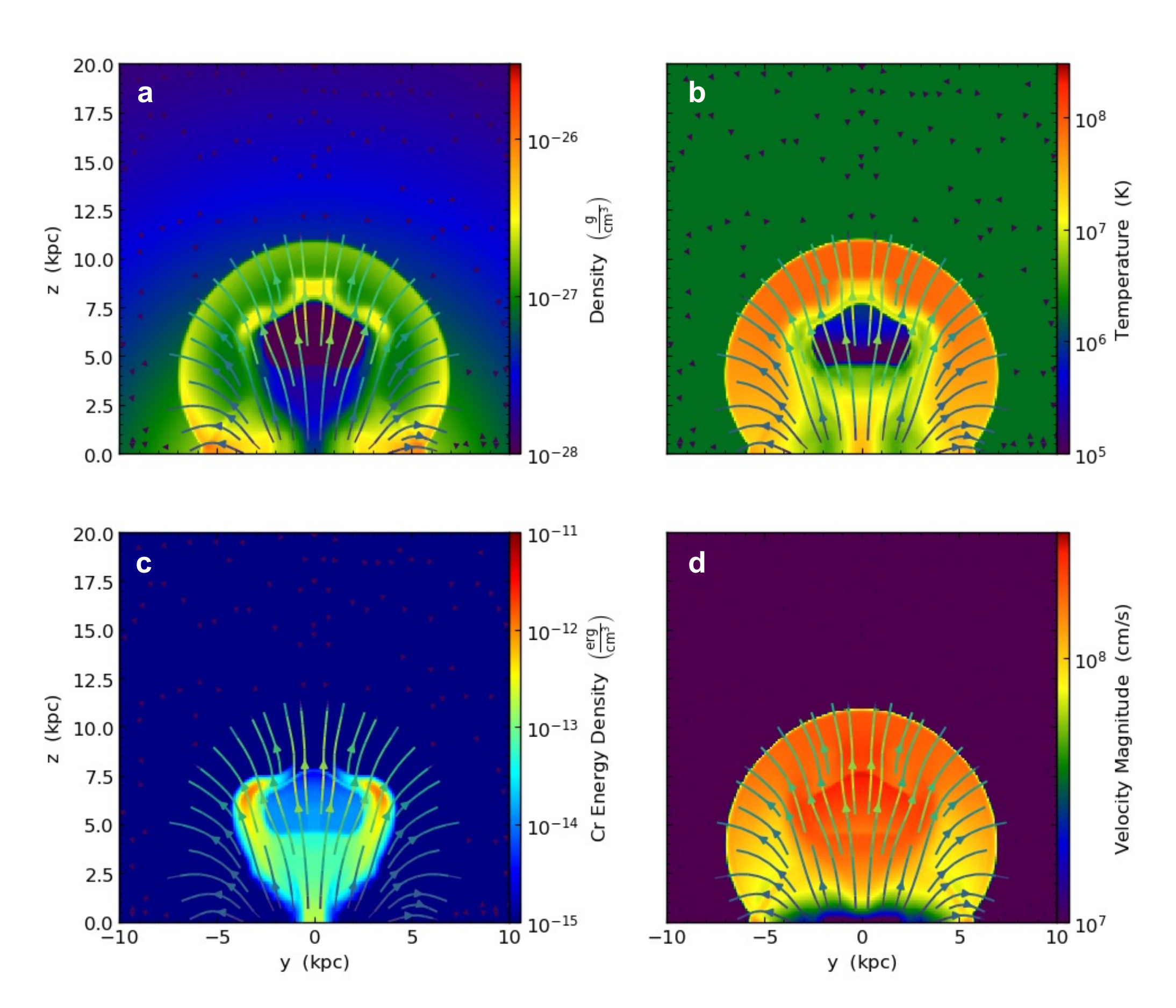}
\caption{{\bf Simulated gas and CR properties.} Slices of the simulated (a) gas density, (b) temperature, (c) CR energy density, and (d) absolute magnitudes of the velocity field at $t=2.6$ Myr. Streamlines show the directions of the outflows driven by the jet injection from the GC.}
\label{fig:slices}
\end{figure}

\begin{figure}[ht]
\centering
\includegraphics[width=\linewidth]{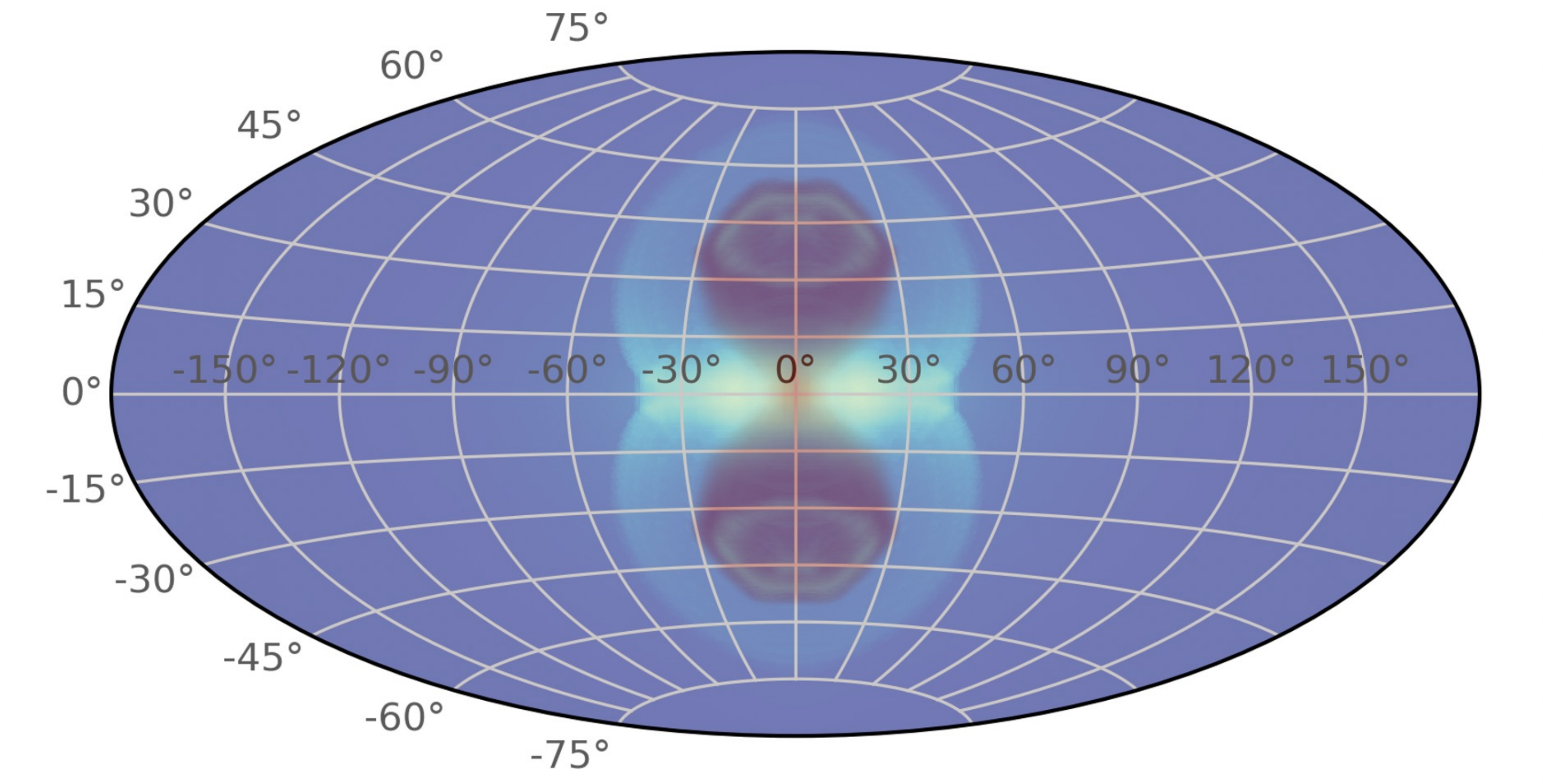}
\caption{{\bf Mock all-sky maps centered at the GC.} The morphology of the simulated gamma-ray (dark purple; 1-2 GeV) and X-ray (blue-yellow; 0.6-1 keV) bubbles resembles the {\it Fermi} bubbles and the {\it eRosita} bubbles, respectively. Our simulations suggest that these fascinating structures in the Milky Way Galaxy can be generated by a powerful outburst from the GC supermassive black hole about 2.6 million years ago.}
\label{fig:map}
\end{figure}

We performed three-dimensional hydrodynamic simulations of energy release at the GC in the form of bipolar jets perpendicular to the Galactic plane. The simulations self-consistently model the evolution of cosmic rays (CRs) injected with the jets at the GC, including their dynamical interactions with the thermal gas within the Galactic halo, and energy losses of CR electrons due to synchrotron radiation and inverse-Compton (IC) scattering as they travel within the Galactic magnetic and radiation fields (see Methods for the simulation details). Modeling of the CRs together with the thermal gas allows us to compute the thermal radiation from the gas and the non-thermal emission generated by the CRs self-consistently. For the simulation results presented in this work, the jet activity of the central SMBH is assumed to start 2.6 Myr ago and last for 0.1 Myr (see Methods for discussion on the chosen parameters). Because of the large pressure contrast with respect to the ambient gas, the jet material expanded into a pair of ``cocoons" or ``bubbles" above and below the GC, similar to radio bubbles observed in galaxy clusters. At the present day, the cocoons have grown and reached a height of $\sim 7.5$ kpc from the Galactic plane (Figure \ref{fig:slices}). The CR electrons within the cocoons that were transported from the GC interact with the interstellar radiation field (ISRF) and shine in the gamma-ray band as the observed {\it Fermi} bubbles extending to Galactic latitudes of $|b| \sim 50^\circ-55^\circ$ (Figure \ref{fig:map}). The same energy injection from the black hole and subsequent cocoon expansion pushed the gas within the Galactic halo away from the GC with supersonic speeds, forming an outward propagating shock. At the shock front, the compression of gas caused an increase in the local gas density, producing enhanced thermal Bremsstrahlung emission in the X-ray band manifested as the {\it eRosita} bubbles. 

\begin{figure}[ht]
\centering
\includegraphics[width=\linewidth]{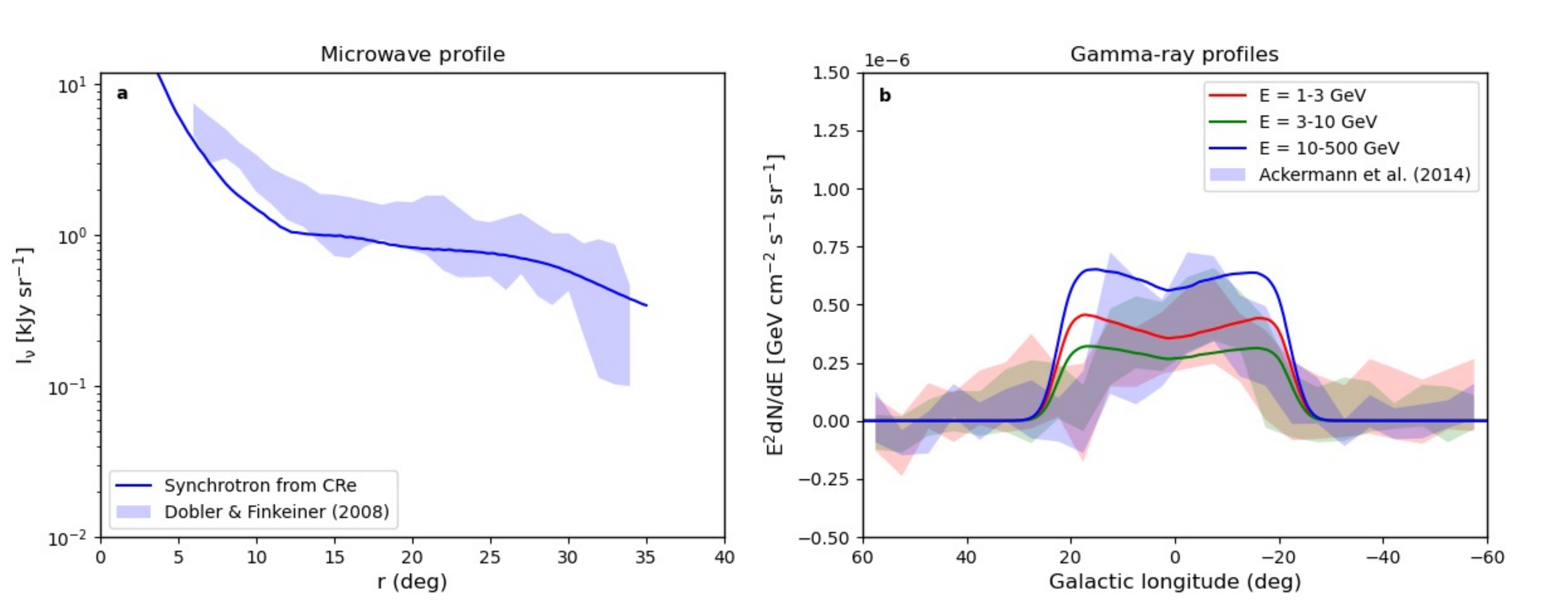}
\caption{{\bf Gamma-ray and microwave profiles.} (a) Intensity profiles of the microwave haze at 23 GHz as a function of radius from the GC. The blue curve shows the predicted synchrotron emission from CR electrons (CRe) from the leptonic jet model, and blue shaded region shows the observed limits of the {\it WMAP} haze\cite{Dobler08}. (b) Gamma-ray intensity profiles ($N$ is the number of gamma-ray photons per unit area per solid angle per unit time) as a function of Galactic longitudes averaged over the latitude region of $40^\circ < |b| < 50^\circ$. Curves show the predicted IC emission, and shaded regions show the observed limits\cite{Ackermann14}. Colors correspond to different energy bins (red: $E=1-3$ GeV; green: $E=3-10$ GeV; blue: $E=10-500$ GeV).}
\label{fig:profiles}
\end{figure}

As can be seen from Figure \ref{fig:map}, the morphology and sizes of both the {\it eRosita} bubbles and {\it Fermi} bubbles are successfully reproduced by the leptonic jet model. Importantly, unique features including their brightness distributions, sharp edges, and smooth surfaces match observational constraints well (see Figure \ref{fig:profiles} and Methods for detailed discussion). For the {\it Fermi} bubbles, reproducing all the above characteristics {\it simultaneously} has proved to be a challenging task for many theoretical models \cite{Yang18}, because they require very specific conditions to be met during the bubble formation. For instance, in the leptonic scenario, the intensity of the ISRF decays with increasing distance from the GC. Consequently, the CR energy density needs to be somewhat enhanced toward higher Galactic latitudes. This enhancement could be caused by initial adiabatic compression when the AGN jets were active \cite{Yang12, Yang17}. The sharp edges of the gamma-ray bubbles require suppression of CR diffusion across the bubble surface, which could be a result of anisotropic CR diffusion along magnetic field lines that are wrapped around the bubble surface due to the magnetic draping effect \cite{Yang12}. Furthermore, the smooth surface of the {\it Fermi} bubbles indicate that there are no vigorous hydrodynamic instabilities on the scale of the bubble sizes. This could be explained by insufficient time for the instabilities to grow \cite{Yang12}, or by suppression of instabilities owing to viscosity \cite{Guo12b}.

% Discussion about NPS (LaRocca 2020: GC, Das 2020: local) 

\begin{figure}[ht]
\centering
\includegraphics[width=\linewidth]{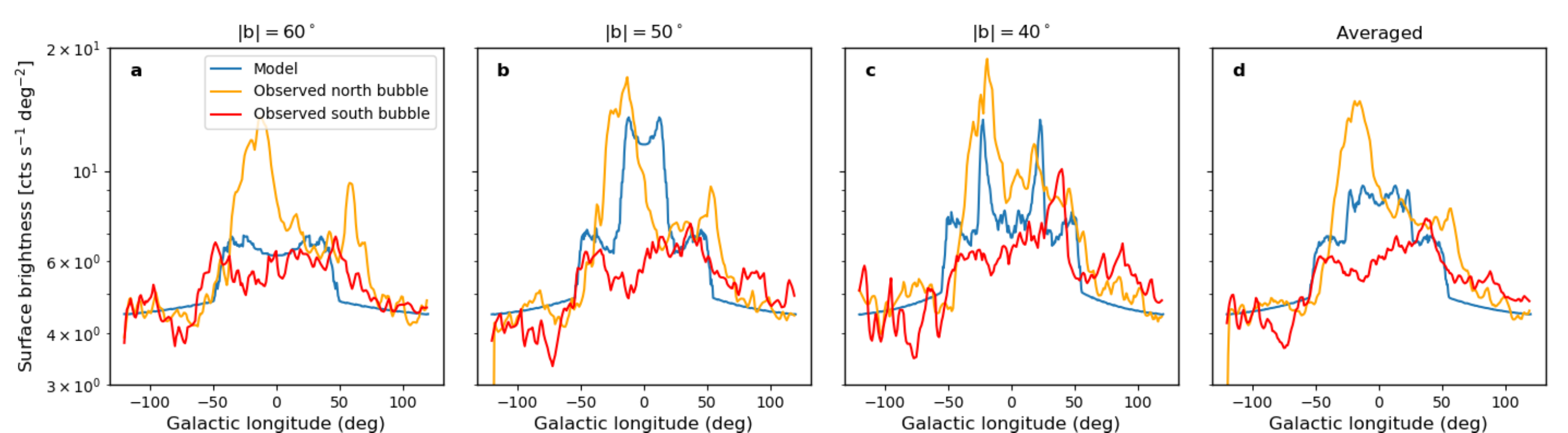}
\caption{{\bf X-ray surface brightness profiles.} Profiles of X-ray surface brightness (in units of counts s$^{-1}$ deg$^{-2}$) as a function of Galactic longitudes in the 0.6-1.0-keV range comparing the model predictions (blue) and the observed profile of the {\it eRosita} bubbles (north and south bubbles shown in orange and red colors, respectively). Panels from left to right correspond to horizontal cuts at Galactic latitudes of $|b|=60^\circ, 50^\circ$, $40^\circ$, and the latitude-averaged profiles. The overall amplitudes and locations of the X-ray shells are well reproduced by the model, though interpretation is complicated by the asymmetry in the data caused by the North Polar Spur observed in the northern X-ray sky.}
\label{fig:xrayprf}
\end{figure}

The X-ray emission predicted by the leptonic jet model shows very good agreement with the observed {\it eRosita} bubbles as well, not only in terms of the extension of the X-ray bubbles, but also in terms of the X-ray surface brightness variations. Figure \ref{fig:xrayprf} shows the comparisons between the simulated and observed X-ray surface brightness profiles at three horizontal cuts, $|b|=40^\circ, 50^\circ$, $60^\circ$, as well as the latitude-averaged profiles. Overall, the predicted amplitudes of the brightness variations and the locations of the X-ray bubble edges are largely consistent with the observational data. In our simulation, the forward shock compresses the gas into a thick shell with width of $\sim 2.5$ kpc (Figure \ref{fig:slices}), consistent with estimates from simple geometrical models \cite{eRosita}. Due to the projection of the compressed gas shell enclosing the gamma-ray bubbles with low gas densities, the modeled X-ray profiles are limb-brightened at the Galactic longitude of $|l|\sim 50^\circ$, similar to the data. At lower latitudes ($|b|=40^\circ$ and $50^\circ$), there is some emission at $|l|\sim 15^\circ-20^\circ$ due to gas that is ejected within the AGN jets being compressed near the contact discontinuity. This feature may be related to the two peaks at $|l|\sim 20^\circ$ for $|b|=40^\circ$ seen in the observed profile for the north X-ray bubble. Despite the large scatter seen in the data, the simulated latitude-averaged profile of the X-ray emission as a function of Galactic longitude approximately falls between the observed values corresponding to the emission from north and south bubbles.
%
%In particular, the limb-brightened surface brightness profiles are recovered, as the forward shock compresses the gas into a thick shell with width of $\sim 2.5$ kpc, consistent with estimates from simple geometrical models \cite{eRosita}. 
%{\bf \color{red} MR: following the report, we decided to replace the latitude-averaged figure with the new one. Could we nevertheless add here the following sentence: ``We note given the large scatter seen in the data, the simulated latitude-averaged profile of the X-ray emission as a function of Galactic longitude (not shown), approximately falls between the observed values corresponding to the emission from north and south bubbles.''}
It is difficult to establish an exact match to the data though, because the observed X-ray sky in this region has many complex features. In particular, the northeast part of the {\it eRosita} bubble is coincident with a prominent structure called North Polar Spur (NPS). The NPS was discovered several decades ago, but its origin remains elusive and is a subject of ongoing debate. Due to the spatial correlation with the Loop I feature in the radio band, some proposed that the NPS is associated with a local superbubble in the Solar neighborhood \cite{Berkhuijsen71, Das20}; other evidence supports scenarios related to past GC activity \cite{Sofue00, Kataoka18, LaRocca20}. The discovery of the south {\it eRosita} bubble has further strengthened the GC hypothesis. Regardless of its origin, the NPS generates a substantial degree of asymmetry in the X-ray sky, complicating the interpretations of the true emission from a GC event. While these considerations prohibit a firm conclusion, our simulated X-ray profiles suggest that an energy outburst from the GC could contribute to the NPS emission, at least partially (i.e., the peak of emission at $l \sim -20^\circ$). This is in line with the conclusion drawn from a recent study of the radio/optical polarization data near the NPS\cite{Panopoulou21}, which suggests that the NPS could be a superposition of both local and GC structures (see Methods for more detailed discussion).

\begin{figure}[ht]
\centering
\includegraphics[width=\linewidth]{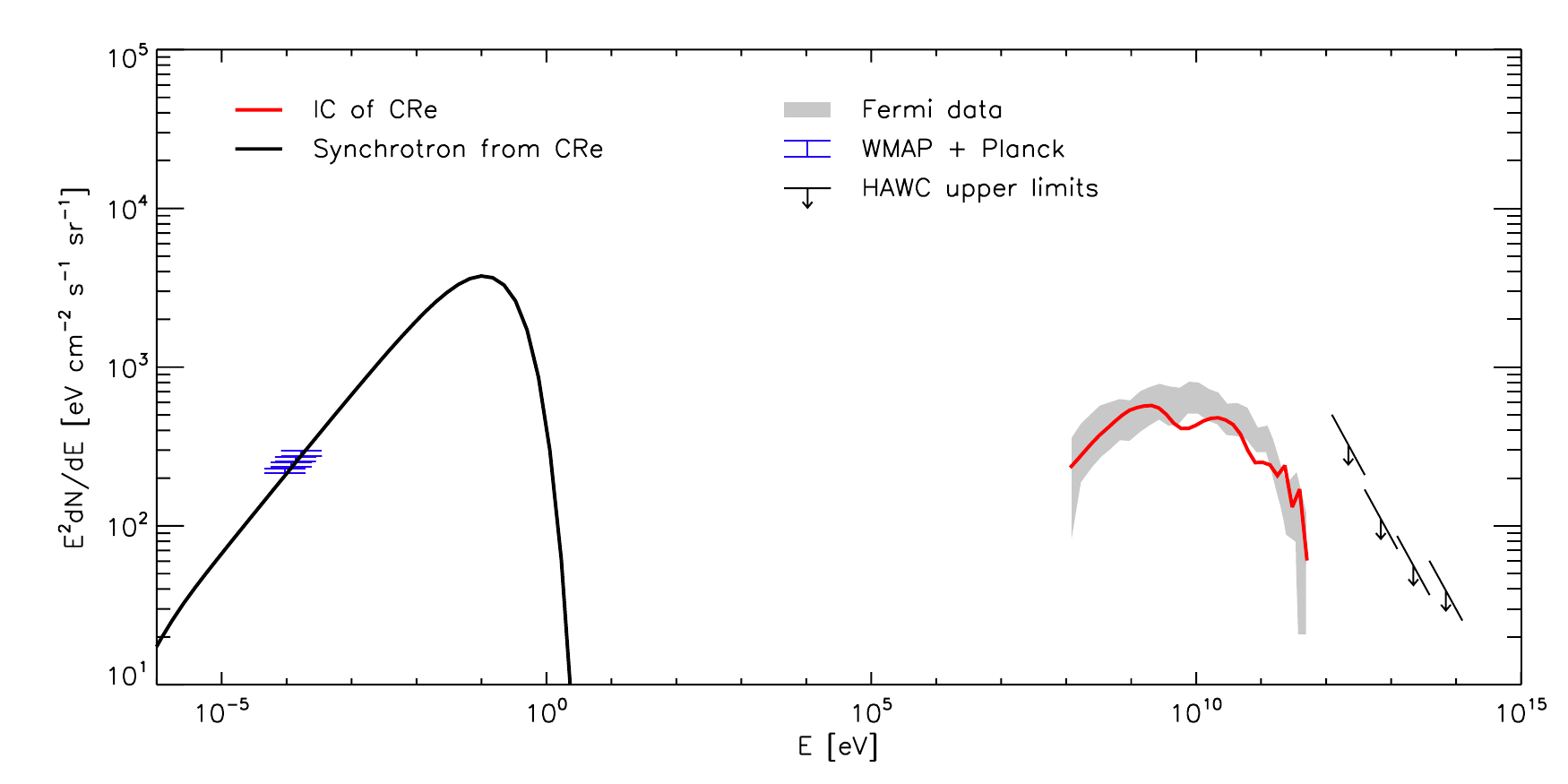}
\caption{{\bf Broad-band spectrum for the Fermi bubbles.} Multi-wavelength spectrum for the region containing the {\it Fermi} bubbles ranging from microwave emission observed by {\it WMAP} and {\it Planck}, GeV gamma rays by {\it Fermi}, and TeV gamma-ray upper limits by {\it HAWC}. For direct comparisons, we computed the simulated spectra within the same extracted regions as the observed spectra, namely, $|l| < 10^\circ, |b|=20^\circ - 40^\circ$ for the gamma-ray bubbles, and $|l| < 25^\circ, |b|=10^\circ - 35^\circ$ for the microwave haze \cite{Ackermann14}. The black and red curves show the predicted emission of the leptonic jet model, in which the primary CR electrons can simultaneously produce the observed gamma-ray emission due to IC scattering and the microwave haze emission by synchrotron radiation, while satisfying the {\it HAWC} constraint.}
\label{fig:spec}
\end{figure}

Additional constraints on the formation of these gigantic bubbles come from the broad-band spectra in the same region in the sky from microwaves by {\it WMAP} and {\it Planck} \cite{Finkbeiner04, PlanckHaze}, GeV gamma rays by Fermi \cite{Su10}, to TeV gamma rays by {\it HAWC} \cite{HAWC17}. We compiled the available observational constraints and our simulated spectra in Figure \ref{fig:spec}. %The {\it Fermi} spectrum is extracted for a region with $|l| < 10^\circ, |b|=20^\circ - 40^\circ$, and the microwave spectrum is for $|l| < 25^\circ, |b|=10^\circ - 35^\circ$ \cite{Ackermann14}. 
In agreement with previous studies \cite{Ackermann14}, we find that the leptonic model can simultaneously reproduce the GeV gamma-ray spectrum of the {\it Fermi} bubbles and the microwave spectrum of the {\it WMAP}/{\it Planck} haze. In the model, the primary CR electrons injected by the AGN jets are quickly transported to large distances before they cool due to synchrotron and IC losses. Therefore, the CR electrons could maintain their hard spectrum (spectral index of -2.1) during propagation, which is necessary for producing the hard gamma-ray and microwave spectra seen in the data. In addition, there appears to be a high-energy cutoff in the {\it Fermi} data at $\sim 110$ GeV, and so far no detection in the TeV range has been made. This high-energy cutoff is also a generic feature predicted by the leptonic jet model. In the early phases of the evolution, the CR electrons suffer great synchrotron and IC losses due to strong magnetic field and radiation field near the GC, generating an exponential cutoff at high energies in their spectrum. After the jets leave the GC, the dynamical timescale of the jets becomes shorter than the synchrotron and IC cooling times, and therefore the CR spectral shape is essentially frozen during the later expansion \cite{Yang17}. Overall, the broad-band spectrum from the bubbles can be well explained by a single population of CR electrons that were transported from the GC by jets from the central SMBH.      

The good agreement between the simulation predictions and the images and spectra of the {\it Fermi} bubbles, {\it eRosita} bubbles, and microwave haze, suggests that past jet activity from the GC black hole could be the common origin of these fascinating structures in our Milky Way Galaxy. The jet parameters adopted in the simulations are summarized in Table \ref{tbl:jet params}. Although these parameters may not be unique (see Methods for discussion), they could still provide us with valuable insights into the formation processes. We find that the new X-ray data from {\it eRosita} allows better constraints on the duration of the jets, as the duration controls the separation between the forward shock (edges of the X-ray bubbles) and the contact discontinuity (edges of the gamma-ray bubbles) in the model. Our simulations suggest that the forward shock at the present day has reached a vertical height of $\sim 11$ kpc away from the Galactic plane and a width of $\sim 14$ kpc, i.e., an oblate ellipsoid rather than a sphere (Figure \ref{fig:slices}). Because the sizes of the {\it eRosita} bubbles are comparable to the distance between the Sun and the GC ($\sim 8.5$ kpc), the top and bottom of the {\it eRosita} bubbles in fact correspond to the near side of the shock front which is only $\sim 6$ kpc away from the Sun roughly in the vertical direction, rather than the top of the shock front near the rotational axis of the Galaxy. In the shock downstream, the electron number density is enhanced to $n_{\rm e}\sim 10^{-3}$ cm$^{-3}$ and the gas is heated to $T\simeq 10^8$ K. At the present day, the forward shock is moving at a speed of $\sim 2000$ km s$^{-1}$ (Mach number $M\sim 10$) in the vertical direction and $\sim 1300$ km s$^{-1}$ ($M\sim 6$) in the lateral direction at a height of 5 kpc away from the Galactic plane. Inside the contact discontinuity, the electron number density is $n_{\rm e}\sim 10^{-5}-10^{-4}$ cm$^{-3}$ and gas temperature is $T \sim 10^5 - 10^7$ K (though the temperature distribution within the contact discontinuity is uncertain due to parameter degeneracies; see Methods). According to our model, the central SMBH was active $\sim 2.6$ Myr ago, injecting a pair of bipolar jets in mostly kinetic forms for a duration of $\sim 0.1$ Myr. After taking into account uncertainties in the initial conditions (Methods), the Sgr A* was estimated to be accreting at $\sim 0.1-1$ the Eddington rate during the active phase, corresponding to a consumption of $\sim 10^3 - 10^4$ solar masses within $\sim 0.1$ Myr.

Our model of the {\it Fermi/eRosita} bubbles makes a specific prediction that could be tested with future observations. The region between the contact discontinuity and the forward shock is compressed to electron number densities of a few $\sim 10^{-3}$cm$^{-3}$ and shock heated to a few $10^7$ K, which are the conditions comparable to those typically observed in the intracluster medium of cool core galaxy clusters. Unlike in most of the Milky Way halo, this region could produce iron K$\alpha$ emission from hydrogen-like (6.7 keV) or helium-like (6.9 keV) iron ions that have peak collisional ionization equilibrium fractions at a few $10^{7}$ K and $10^{8}$ K, respectively\cite{Ezoe2021}. 
We note that whether such a line is detectable depends on the fraction of the fast hot gas. Since the temperature of the gas is uncertain in our model, that fraction may be too small for existing instruments to detect. However, it may be detectable with {\it Athena} as its effective area is almost two orders of magnitude larger than {\it Chandra}’s at 6.9 keV \cite{Barret2020}.
Given the expected 
%Mach number $\sim 20$, and the 
post-shock velocity of $\sim 2000$ km s$^{-1}$ (see lower right panel in Figure \ref{fig:slices}), 
%the post shock region should move at $\sim 500$ km s$^{-1}$, which corresponds to 
the total Doppler line broadening is expected to be about 45 eV at 6.7 keV line energy, but the exact value will depend on the details of the velocity projection. 
A shift of this magnitude should be detectable with the {\it Athena} X-ray Integral Field Unit that has a planned spectral resolution of 2.5 eV up to line energies of 7 keV and more than sufficient spatial resolution to probe this region \cite{Barret2018}. 
Thus, a coherent outflow originating from the region between the shock and contact discontinuity could be a testable model prediction, especially if the data quality is sufficient to constrain not only the line width but also its skewness or profile.

% SEE IF WE COULD OBTAIN ETH WITHIN THE EROSITA BUBBLES

% Comparisons with the observed features of eRosita bubbles: sharpness, wide bottom, parameters (e.g., Mach number, brightness, kT, t_cool, v_exp, age, Eth~1e56 erg in Predehl 2021, real size ~ 5-6 kpc, roughly constant pressure within the eRosita bubble)

\section*{Discussion}

% leptonic jet model vs. in situ acceleration models (Sarkar 2015, 2017, 2019)

As mentioned in the Introduction, both the leptonic jet model and the model of in situ acceleration by shocks or turbulence are promising mechanisms to explain the gamma-ray emission of the {\it Fermi} bubbles, but new insights could be obtained by adding the constraints from the {\it eRosita} data. One primary difference between the leptonic jet model and the shock acceleration model is the location of the forward shock driven by the outburst. In the leptonic jet model, the surface of the {\it eRosita} and {\it Fermi} bubbles corresponds to the forward shock and contact discontinuity, respectively. On the other hand, if the {\it Fermi} bubbles are produced by shock accelerated CRs as proposed in some of the in situ acceleration models\cite{Zhang20}, the more extended {\it eRosita} bubbles would be left unexplained. In this regard, the leptonic jet model may be favored as both the {\it eRosita} bubbles and the {\it Fermi} bubbles could be simultaneously accounted for by a single event. 
In situ acceleration models in which the {\it Fermi} bubbles are produced by CRs accelerated by turbulence generated within the bubbles\cite{Mertsch19} remain a possibility. In this scenario, the forward shock driven by the outflow could account for the {\it eRosita} bubbles, and the turbulence in the shock downstream could potentially stochastically accelerate CRs that are responsible for the gamma-ray bubbles. However, note that the surface of the {\it Fermi} bubbles is very smooth, suggesting that hydrodynamic instabilities at the bubble surface are suppressed, and hence the amount of turbulence generated may be limited. Also, in this scenario the turbulence is expected to be volume-filling in the shock downstream and of increasing strength toward the shock front \cite{Yang13}, and therefore the sharp edges of the {\it Fermi} bubbles cannot be naturally explained. More detailed simulations are needed to test whether the stochastic acceleration models could satisfy all the observational constraints. 

% Discussion about pros and cons of the leptonic jet model, e.g., ionization cone in MS (0.1-1 L_edd, age~3 Myr, Bland-Hawthorn 2019), gas kinematics (Fox 2015, Bordoloi 2017, Ashley 2020), OVIII/OVII ratios (logT~6.6, age~3-4Myr, L_mech~2e42 erg/s, Miller & Bregman 2016, Sarkar 2017)

Our simulation suggests that a single past jet activity $\sim 2.6$ Myr ago from the GC black hole could be the common origin for the {\it eRosita} bubbles, the {\it Fermi} bubbles, and the microwave haze. The required Eddington ratio is estimated to be $\sim 0.1-1$ during the brief active accretion phase of $\sim 0.1$ Myr. Though such high activity of Sgr A* may appear somewhat surprising given its extreme quiescence at the present day, multiple lines of evidence have pointed to much elevated past activity of Sgr A* over the past $\sim 10$ Myr\cite{Totani06}. In particular, recent studies have found enhanced ionization levels in the Magellanic Stream, which could be explained if Sgr A* went through a phase of Seyfert-like flaring activity a few Myr ago\cite{BlandHawthorn19}. The energetics and timescales obtained by our simulations are in good agreement with such a Seyfert-flare scenario. The epoch of outburst predicted by our model is also in line with several independent constraints from kinematic studies of the halo gas using X-ray and UV absorption lines\cite{Fox15, Miller16, Bordoloi17}, which estimated that the outburst occurred $\sim 2-8$ Myr ago (though there remain uncertainties in the interpretations; see Methods for detailed discussion).
%\textcolor{blue}{The phrasing here somewhat confuses me. I would have used ``timescale" to mean the 0.1 Myr duration of jet activity, and ``history" or ``epoch" to refer to the 2-8 Myr period.  }

% Discussion open questions and future prospects (e.g., how to distinguish starburst and AGN)

The rich multi-wavelength observational data as well as detailed theoretical modeling have provided valuable information about the past GC activity. For example, our model suggests that the radiation and magnetic fields are likely suppressed in the inner kpc region near the GC at the time when the jets were launched (see Methods). Such suppression could perhaps be related to an earlier injection that produced the GC chimney\cite{Ponti19} and bipolar radio bubbles\cite{Heywood19} recently observed close to the Galactic plane. Alternatively, the suppression could be associated with the formation of the nuclear star cluster $\sim 6$ Myr ago\cite{Paumard06}, where stellar winds or supernova explosions of the massive young stars could have helped to evacuate the gas and create a GC environment less hostile to the cooling CR electrons within the jets. Such a scenario is also broadly consistent with the emerging picture of SMBH-galaxy co-evolution in that AGN activity often accompanies star formation activity in the galaxy\cite{Heckman14}. Some open questions remain to be addressed, e.g., whether CRs accelerated at the forward shock could generate observable non-thermal emission in the leptonic jet scenario, whether the thermal structure of the Galactic halo probed by X-ray absorption line studies\cite{Miller16} could be explained, how the jet-induced outflows entrain cold gas and how it may be related to the high velocity clouds observed in the Galactic halo\cite{Ashley20}. Future investigations will further reveal the impact of this energetic feedback on the evolution history of our Milky Way Galaxy, and how this event fits in the broader picture of SMBH-galaxy co-evolution in the universe.

\newpage

\section*{Methods}
\label{sec:Methods}

\subsection*{Simulation setup and parameters}

% Methods -- summary and key differences compared to previous simulations

We carried out three-dimensional hydrodynamic simulations of bipolar jets emanating from the GC and perpendicular to the Galactic plane including relevant CR physics using the FLASH code \cite{Flash,Lee09,Yang12}. We utilized the CRSPEC code to self-consistently model the evolution of CR spectrum due to synchrotron and IC cooling, while accounting for the dynamical interaction between the CRs and the thermal gas \cite{Yang17}. The modeling of the CR spectral evolution is crucial, as it allows us to simultaneously compute the non-thermal radiation from the CRs and the Bremsstrahlung emission from the thermal gas. The injected CR electrons are assumed to follow an initial power-law spectrum ranging from 10 GeV to 10 TeV, with a spectral index of -2.1. In our previous work \cite{Yang13}, we demonstrated that the magnetic field within the {\it Fermi} bubbles needs to be amplified to values comparable to the ambient field at the present day in order to reproduce the microwave haze emission. Therefore, we do not simulate the magnetic field directly, but adopt the default magnetic field distribution in GALPROP \cite{Strong09} for the computation of the microwave haze, i.e., $|B|=B_0\exp(-z/z_0)exp(-R/R_0)$, where $R$ is the projected radius to the Galaxy's rotational axis, $B_0$ is the average field strength at the GC, and $z_0$ and $R_0$ are the characteristic scales in the vertical and radial directions, respectively. We adopt $z_0=2$ kpc, $R_0=10$ kpc, and $B_0=50\mu$G as motivated by observational constraints\cite{Crocker10}. As will be discussed later, the high-energy cutoff in the observed gamma-ray spectrum requires suppression of radiation and/or magnetic field strengths near the GC at the time of jet injection. Therefore, for computing the synchrotron losses in the simulations, we have used a factor of three lower normalization of magnetic field strength compared to the value adopted in our earlier work\cite{Yang17}. For IC scattering, we utilize the ISRF model in GALPROP and compute the CR electron energy losses and gamma-ray emissivity including the Klein-Nishina effect. To generate the X-ray image, the X-ray emissivity in the energy range 0.6-1.0 keV is calculated using the MEKAL model implemented in XSPEC \cite{Arnaud96}. The initial condition for the halo gas is the same as in previous leptonic jet models\cite{Guo12a, Yang12}, which assumes an isothermal halo with $T=2\times 10^6$ K in hydrostatic equilibrium within a fixed Galactic potential. The normalization of the gas density profile is chosen to match the observed profile derived from X-ray absorption line studies\cite{Miller13}. Other simulation setups are essentially identical to our previous works \cite{Yang12, Yang13, Yang17}, though the new {\it eRosita} data motivated some parameter adjustments. We summarize the parameters in Table \ref{tbl:jet params} and discuss key differences as follows. 

In previous studies of the leptonic jet model \cite{Guo12a, Yang12}, it was found that there are parameter degeneracies when the morphology of the gamma-ray bubbles was the primary constraint. The new {\it eRosita} data provides another key constraint on the separation between the forward shock and the contact discontinuity in the model. Compared to the parameters adopted in our previous simulations \cite{Yang17}, we find that the duration of the jets has to be reduced from $t_{\rm jet} = 0.3$ Myr to $t_{\rm jet} = 0.1$ Myr. Changing the jet duration alone would produce {\it Fermi} bubbles that are too oblate. In order to maintain the right axial ratio of the {\it Fermi} bubbles, we then need to decrease the total (thermal plus CR) injected energy density to one third of the original value, i.e., from $e_{\rm j} + e_{\rm jcr} = 4.09 \times 10^{-9}$ to $1.36 \times 10^{-9}$ erg cm$^{-3}$. By changing these two parameters, the morphology of both the X-ray and gamma-ray bubbles can be reproduced. Note, however, that the overall dynamics and bubble shapes are controlled by the total injected energy density (thus the total pressure within the bubbles), and hence the individual values of $e_{\rm j}$ and $e_{\rm jcr}$ are degenerate in the current model. Due to this parameter degeneracy, the predicted gas temperature within the {\it Fermi} bubbles should be considered uncertain as it depends on the amount of injected thermal energy assumed in the model. Additional observational constraints such as X-ray, UV, and radio emission/absorption lines\cite{Miller16, Bordoloi17, DiTeodoro18, Fox20} may be used to break the parameter degeneracy. We will investigate this issue in future studies.  

The simulations are analyzed using the software {\it yt} \cite{Turk11}, and the simulated gas and CR distributions at $t=2.6$ Myr are shown in Figure \ref{fig:slices}. At this time, the forward shock driven by the AGN outburst has propagated to a vertical height of $\sim 11$ kpc and a radius of $\sim 7$ kpc from the rotational axis of the Galaxy. The shock front is currently expanding with a speed of $\sim 1000-2000$ km s$^{-1}$, with nearly vertical flows near the axis and progressively larger opening angles away from the axis. In the shock downstream, the gas is compressed to $n_{\rm e} \sim 10^{-3}$ cm$^{-3}$ and heated to $T\sim 10^8$ K, producing the X-ray bubbles as observed. Within the contact discontinuity is the underdense cavity filled with thermal gas and CRs, though their relative contributions are uncertain due to the parameter degeneracy discussed above. The CR distribution is edge-brightened, which is necessary for reproducing the nearly flat intensity of the {\it Fermi} bubbles \cite{Yang17}. Similar to our previous findings, we find that only a small fraction ($f_{\rm e} \sim 10^{-2}$) of the simulated CRs needs to be CR electrons in order to match the observed gamma-ray and microwave spectra. Note that the value of $f_{\rm e}$ is larger than that found in our previous studies because of the reduced amount of CR energy injected in the current simulations. 

For completeness, we show the comparisons between the predicted and observed profiles for the microwave haze and gamma-ray bubbles in Figure \ref{fig:map}. As can be seen, the leptonic jet model not only can simultaneously reproduce the spectra of the gamma-ray bubbles and the microwave haze (Figure \ref{fig:profiles}), but also the key features in their spatial distributions. The predicted synchrotron emission follows a centrally peaked profile, primarily because of the exponential decay of the Galactic magnetic field from the GC. As for the gamma-ray profiles, the nearly flat intensity profiles and sharp edges of the observed bubbles are reproduced, which requires very specific three-dimensional distributions of CRs as well as suppression of CR diffusion across the bubble surface (see discussion in the main text). Note, though, that due to large levels of noise and asymmetries in the data, it is impossible to provide a perfect fit to the data. Therefore, we do not aim to complicate the models by introducing additional free parameters to improve the fits as it could result in overinterpretation of the data. However, should future observations be characterized by lower noise, such improvements could be warranted. For instance, the mismatch at $l\sim 20^\circ$ between the predicted and observed gamma-ray profiles may be related to the bending of the observed {\it Fermi} bubbles toward the west and that could be accounted for in the model by invoking a slight jet tilt as considered in our earlier work\cite{Yang12}. As a side note, we point out that in our previous analysis based on magnetohydrodynamic simulations\cite{Yang13}, the CRs needed to be replenished by a second jet injection in order to match the microwave profile. That was because the initial magnetic field strength in the simulation was set to its present-day value, and hence the strong magnetic pressure near the GC pushed the gas and CRs away and caused a depression in the microwave profile at $r\lesssim 15^\circ$. However, our current analysis, which is based on hydrodynamic simulations that do not model the effects of a dynamically important initial magnetic field, shows that a single jet could also yield consistent results with the observed haze profile. Since the second jet scenario\cite{Su12} was no longer supported by the updated {\it Fermi} data\cite{Ackermann14}, this would imply that a single jet plus an initially suppressed magnetic field is the more likely scenario. As we will show later, the constraints from the high-energy cutoff in the gamma-ray spectrum also point to the same conclusion.

%\subsection*{Comparisons with other observational constraints on the Galactic halo}
\subsection*{Comparisons with other observational constraints}
Our model predicts the existence of a thick shell of ultrahot gas ($T\sim 10^8$ K) moving with high velocities ($\sim 1000-2000$ km s$^{-1}$) within the Galactic halo. Is this consistent with existing X-ray observations of the Galactic halo and kinematics of the halo gas as traced by X-ray/UV absorption lines from background quasars? While the outflow velocities and timescales of outburst predicted in our model are consistent with some of the observational studies\cite{Fox15, Miller16, Bordoloi17}, other studies have inferred more mild outflow velocities of $\sim 200-300$ km s$^{-1}$\cite{Kataoka13, Fang14, Sarkar17}. The apparent discrepancies could be explained by taking into account a number of factors: (1) the predicted hot gas in the range of $T > 10^8$ K would not contribute substantial emission in the X-ray band probed by current data. In addition, for strong shocks, it is likely that the ions are preferentially heated to higher temperatures than the radiating electrons. Therefore, outflow velocities inferred by measuring the temperature contrast (where the temperature is obtained by fitting the X-ray spectrum emitted by the electrons) between the shocked gas and the ambient medium may underestimate the true velocities\cite{Kataoka13}; %\textcolor{blue}{I'm not sure I understand the meaning of this. The emission is excited by electrons (generally), but I think the Doppler velocities inferred from it would correspond to the ions. Is this consistent with what you said?} 
(2) the UV absorption lines probe the kinematics of the cooler $T\sim 10^4-10^5$ K gas, and thus using them to infer the velocity of the hot gas would depend on how exactly the cooler gas is formed/entrained in the hot flow, which remains a major unresolved question itself. Furthermore, in an entrainment scenario, the velocities of the cold clouds are typically smaller than those of the hot gas due to imperfect momentum transfer. The inferred velocities of $\sim 1000-1300$ km s$^{-1}$ from UV absorption lines\cite{Fox15, Bordoloi17} would thus imply even higher velocities for the hot flow; and (3) there remain large uncertainties in the interpretations of the observational data, e.g., assumptions about the geometry of the outflows and injection patterns (continuous vs. instantaneous injections), the asymmetric emission measure of the Galactic halo\cite{Kataoka15}, and confusion due to foreground/background projections (e.g., the NPS). Given all the above considerations, we leave the detailed comparisons with these observational data to dedicated future work.

%\subsection*{Constraints on the initial GC environments and particle acceleration}
\subsection*{Initial GC environment and particle acceleration}

% Discussion about constraints on Emax,0 and possible acceleration of particles

As discussed in our previous study\cite{Yang17}, the latitude-independent high-energy cutoff in the observed gamma-ray spectrum at $\sim 110$ GeV implies a very uniform distribution for the maximum CR energy, $E_{\rm max} \sim 300$ GeV. The maximum energy of CR electrons today is set by the fast synchrotron and IC cooling near the GC at early times, followed by adiabatic cooling due to the bubble expansion. These considerations allow us to infer conditions within $\sim$ kpc from the GC in the early phase of the evolution. Because of the parameter modifications mentioned above, the expansion is somewhat slower, and the age of the bubbles becomes somewhat longer than our previous estimates, changing from 1.2 Myr to 2.6 Myr. This leads to longer times for the CR electrons to cool adiabatically, changing $E_{\rm max}$ by a factor of 10 from the simulation time $t\sim 0.4$ Myr to $t=2.6$ Myr. This implies that the maximum CR energy is $\sim 3$ TeV at the time when the jets leave the inner kpc region. According to the constraints we obtained in the previous study\cite{Yang17} (see their Figure 6), it implies that either the initial jet velocity needs to be larger, or the initial radiation plus magnetic field energy density near the GC needs to be suppressed when the jets were first launched. Assuming the initial jet velocity is unchanged (otherwise it would modify other properties such as bubble morphology), the radiation plus magnetic field energy densities inferred would need to be suppressed to $u_{\rm tot} = u_{\rm B} + u_{\rm rad} \sim 1.9\times 10^{-12}$ erg cm$^{-3}$. Assuming the energy density of the ISRF at the GC is $u_{\rm rad} \sim 1.54\times 10^{-12}$ erg cm$^{-3}$, the same as the GALPROP model at the present day after taking into account the Klein-Nishina effect, then the magnetic field strength at the GC at the time of injection would need to be $\sim 3\ \mu$G. Equivalently, if one would like to estimate characteristic CR cooling times, one could define an effective magnetic field strength such that $u_{\rm tot} \equiv |B_{\rm eff}|^2 / (8\pi)$. Our constraints would imply $|B_{\rm eff}| \sim 7 \mu$G.  

One could ask whether the inferred GC conditions are consistent with CR electrons being accelerated up to $\sim 3$ TeV by checking two criteria. First, assuming the CRs within the jets are accelerated by diffuse shock acceleration (DSA), the maximum CR electron energy can be estimated by balancing the DSA acceleration rate and the IC plus synchrotron cooling rate\cite{Rosswog11},
\[
E_{\rm e,max} \sim 6\times 10^4 \xi^{1/2} \left(\frac{B_{\rm eff}}{1 \mu {\rm G}} \right)^{-1/2} {\rm TeV},
\]
where $\xi \sim 0.1$ is the DSA acceleration efficiency. %\textcolor{blue}{This efficiency factor is rather low. Is there any reason not to adopt the canonical 0.1?}
One can see that, given the suppressed radiation and magnetic field strengths, the acceleration of CR electrons is not strongly limited by the synchrotron and IC cooling. Of course, the magnetic field strength within the jets is uncertain and could be higher, especially close to the black hole\cite{Blandford19}. But the above estimate suggests that the condition of $E_{\rm e,max} > 3$ TeV could be satisfied for a wide range of magnetic field strength. The second criterion is the constraint on acceleration timescales. The maximum rate of particle acceleration by a strong parallel shock can be written in the following form\cite{Lagage83, Zweibel03b},
\[
\frac{dE}{dt} = 1.5\times 10^{-24} Z\left(\frac{B}{1\mu{\rm G}} \right) V_{\rm sh}^2 {\rm GeV s^{-1}},
\]
where $Z$ is the charge, and $V_{\rm sh}$ is the shock velocity in cm s$^{-1}$. For $V_{\rm sh}=0.025c$, $B \sim 3\mu$G, and $Z=1$, $dE/dt \sim 8.44\times 10^{-7}$ GeV s$^{-1}$. One could then estimate the required acceleration timescale $t_{\rm acc}=E_{\rm max}/(dE/dt)$ to accelerate CRs to $E_{\rm max} = 3$ TeV, which is much less than the dynamical time of the jets of $t_{\rm dyn} \sim (1\ {\rm kpc})/(0.025c) \sim 0.13$ Myr. In other words, there is sufficient time for CR electrons to be accelerated to $\sim 3$ TeV. Therefore, overall we find that DSA is a plausible mechanism to accelerate CR electrons to the energy required in the model.

Relativistic magnetic reconnection (RMR\cite{SironiSpitkovsky2014}) is also a possible primary particle acceleration mechanism. The properties of RMR depend strongly on the magnetization parameter $\sigma\equiv B^2/(4\pi\rho c^2)$ (here, $\rho$ is the
rest mass particle energy density), with the relativistic regime defined by $\sigma > 1$. It is found that as $\sigma$ increases, the energetic particle spectral index $p$ (assuming the particle energy spectrum is a power law distribution, $E^{-p}$) decreases, and can drop below 
 $p = 2$. This raises the interesting possibility that the observed $E^{-2}$ spectrum is actually an aged spectrum, which would allow the jet to be older and/or the particle transport slower than what is inferred from DSA. However, the hardening of the spectrum is accompanied by a decrease in the maximum particle energy achievable (as it must be): $\gamma_{\rm max}\sim [(\sigma + 1)(2-p)/(p-1)]^{1/(2-p)}$. Thus, to achieve 3 TeV ($\gamma_{\rm max}\sim 6\times 10^{6}$), either $\sigma$ must be very large or $p$ must be very close to 2. For example, if $p=1.5$, $\sigma\sim 2.4\times 10^3$, while if $p=1.9$, $\sigma$ could be as small as $42$. But with $p$ so close to 2, the constraints on aging are essentially the same as for DSA. Thus, we regard the constraints on jet age and particle transport as robust.

\subsection*{Uncertainties and limitations of the current model}

% Discussion about the initial density profile and how the jet power should be an upper limit due to cuspy initial density profile assumed. 2nd argument: B suppression from 50 to 3 muG, and B \propto \rho^(2/3).

The power of the two jets obtained in our model is $6.32\times 10^{44}$ erg s$^{-1}$, which is close to the Eddington luminosity for Sgr A*. Note, however, that the estimated jet power in our model is directly proportional to the unknown initial central gas density assumed in the simulations\cite{Yang12}. Although we have used the observational constraints from X-ray absorption line studies\cite{Miller13} to inform the gas profile of the Galactic halo, the true value near the GC at the time of injection remains highly uncertain. As discussed in the main text, if the gas near the GC were evacuated due to a prior energy injection or the formation of the nuclear star cluster, it is conceivable that the central density 
%may be 
is lower than that assumed in our model. Therefore, the energetics of the SMBH accretion event obtained in our model shall be considered as an overestimate, and the Eddington ratio is likely to be in the range of $\sim 0.1-1$ given the above considerations.  

Our simulation setup of the Galactic halo is relatively simple and is strictly symmetric about the GC. In reality, the Galactic halo is much more complex. The most prominent structure is the NPS, which is very bright in the north-eastern X-ray sky. As mentioned in the main text, it remains uncertain whether the NPS is associated with a local superbubble or a GC event. Generally speaking, studies based stellar polarization and extinction tend to support the local bubble scenario\cite{Berkhuijsen71, Das20, Panopoulou21}, whereas analyses on the basis of X-ray data tend to suggest a GC origin\cite{Sofue00, Kataoka18, LaRocca20}. In order to reconcile the apparent discrepancies, it has been proposed that the X-ray emission of the NPS may be a superposition of both the local and GC structures\cite{Panopoulou21}. Our simulation also supports this conclusion. It has also been reported that the emission measure of the Galactic halo is asymmetric about the Galactic plane\cite{Kataoka15}. If the outflow from the GC is expanding into this asymmetric Galactic medium, the brighter emission of the NPS could also be accounted for\cite{Sarkar19}. Another limitation of our simulation is that we have only modeled the hot component of the halo gas and neglected the colder, multi-phase Galactic disk. Therefore, our simulation results cannot be directly compared to observed structures close to the Galactic plane, including the X-ray shells observed by ROSAT\cite{BlandHawthorn03}, and the more recently discovered HI outflows\cite{DiTeodoro18}, X-ray chimney\cite{Ponti19} and bipolar radio bubbles\cite{Heywood19}. Since all these structures may provide important clues to the formation of the gamma-ray/X-ray bubbles, we will extend our model to include the Galactic disk component and investigate the detailed jet-disk interactions in future work.

In order to inflate the nearly symmetric {\it eRosita} and {\it Fermi} bubbles, our model requires bipolar jets in the direction perpendicular to the Galactic plane. Generally speaking, the directions of SMBH jets are determined by the orientation of accretion disks and/or the black hole spins on much smaller scales and do not need to align with the rotational axis of the host galaxies. Future studies would be required to see whether this condition could be relaxed by considering tilted jets interacting with dense, multiphase interstellar medium within the Galactic disk\cite{Cecil21}. Alternatively, such alignment may indeed be expected in low-mass, disky galaxies like the Milky Way, as shown by recent simulations\cite{Fiacconi18}. In this case, the rotational axis of the black hole accretion disk may align with that of the host galaxy as it is fed by gas with high angular momentum, and the black hole spin could efficiently align with the accretion disk due to the Bardeen-Petterson effect\cite{Bardeen75} because of the relatively small mass of the black hole compared to its accretion disk.

% Discussion about parameter variations and the caveat that this parameter set may not be unique.

Finally, we note that the computational costs of our three-dimensional, CR hydrodynamic simulations forbid us to do a full parameter scan of all the possible combinations of the six jet parameters. Therefore, the jet parameters we found in this study may not be unique, even though the parameter space should be very limited given the stringent constraints from all the available observational data\cite{Yang12}. Nevertheless, our current simulation serves as a proof of concept that the gigantic bubbles within our Galaxy could plausibly originate from past jet activity of the GC black hole. 

\subsection*{Data Availability}
Simulation data that were used to generate the emissivity profiles are available in the Supplementary Information. Source data associated with other main figures of the manuscript are available from the corresponding author upon reasonable request. 

%Simulation data that support the findings of this study or were used to make the plots are available from the corresponding author upon reasonable request. Source data associated with the main figures of the manuscript are available from the corresponding author upon reasonable request. Source data associated with the main figures of the manuscript are available at

%The datasets generated during and/or analysed during the current study are available from the corresponding author on reasonable request.

\subsection*{Code Availability}
The simulations were performed using the code FLASH, publicly available at https://flash.rochester.edu/site/flashcode/, with modifications described in refs. 13 and 15. The CR module is a proprietary software product funded by NASA and NSF and is not publicly available.

%The custom code utilized for the current study are available from the corresponding author to editors and reviewers upon request.

\newpage

\section*{Acknowledgements}

%Acknowledgements should be brief, and should not include thanks to anonymous referees and editors, or effusive comments. Grant or contribution numbers may be acknowledged.

H.Y.K.Y. acknowledges support from Yushan Scholar Program of the Ministry of Education of Taiwan and Ministry of Science and Technology of Taiwan (MOST 109-2112-M-007-037-MY3). M.R. acknowledges support from NSF Collaborative Research Grants AST-1715140 and AST-2009227, 
and NASA grants 80NSSC20K1541 and 80NSSC20K1583. E.G.Z. acknowledges support from NSF Collaborative Research Grant AST-2009323. The simulations are performed and analyzed using computing facilities operated by the National Center for High-performance Computing and the Center for Informatics and Computation in Astronomy at National Tsing Hua University. FLASH was developed largely by the DOE-supported ASC/Alliances Center for Astrophysical Thermonuclear Flashes at University of Chicago. Data analysis presented in this paper was conducted with the publicly available {\it yt} visualization software\cite{Turk11}. 

\section*{Author contributions statement}

%Must include all authors, identified by initials, for example:
%A.A. conceived the experiment(s),  A.A. and B.A. conducted the experiment(s), C.A. and D.A. analysed the results.  All authors reviewed the manuscript. 

H.Y.K.Y. carried out the simulations and analyses and prepared the manuscript. M.R. participated in the interpretation of the simulation results and assisted in the preparation of the manuscript. E.G.Z. contributed to the discussions of particle acceleration and assisted in the preparation of the manuscript.

\section*{Competing interests}
The authors declare no competing interests.

\section*{Additional information}

Correspondence and requests for materials should be addressed to H.Y.K.Y. Reprints and permissions information is available at www.nature.com/reprints.

%To include, in this order: \textbf{Accession codes} (where applicable); \textbf{Competing interests} (mandatory statement). 

\newpage

\section*{Figure legends/captions}

% Figure 1 -- Slices of gas density, CR energy density, Emax, etc

% caption only
%{\bf Simulated gas and CR properties.} Slices of the simulated (a) gas density, (b) temperature, (c) CR energy density, and (d) absolute magnitudes of the velocity field at $t=2.6$ Myr. Streamlines show the directions of the outflows driven by the jet injection from the GC.

\newpage

% Figure 2 -- Mock all-sky map of the simulated gamma-ray and X-ray emission, similar to Fig. 3 of the eRosita paper

% caption only
%{\bf Mock all-sky maps centered at the GC.} The morphology of the simulated gamma-ray (dark purple; 1-2 GeV) and X-ray (blue-yellow; 0.6-1 keV) bubbles resembles the {\it Fermi} bubbles and the {\it eRosita} bubbles, respectively. Our simulations suggest that these fascinating structures in the Milky Way Galaxy can be generated by a powerful outburst from the GC supermassive black hole about 2.6 million years ago.

\newpage

% Figure 3 -- Gamma-ray and microwave profiles

% caption only
%{\bf Gamma-ray and microwave profiles.} (a) Intensity profiles of the microwave haze at 23 GHz as a function of radius from the GC. Blue curve shows the predicted synchrotron emission from the leptonic jet model, and blue shaded region shows the observed limits of the {\it WMAP} haze\cite{Dobler08}. (b) Gamma-ray intensity profiles as a function of Galactic longitudes averaged over the latitude region of $40^\circ < |b| < 50^\circ$. Curves show the predicted IC emission, and shaded regions show the observed limits\cite{Ackermann14}. Colors correspond to different energy bins (red: $E=1-3$ GeV; green: $E=3-10$ GeV; blue: $E=10-500$ GeV).

\newpage

% Figure 4 -- Horizontal profiles of the simulated gamma-ray and X-ray emission at different Galactic latitudes

% caption only
%{\bf X-ray surface brightness profiles.} Profiles of X-ray surface brightness (in units of counts s$^{-1}$ deg$^{-2}$) as a function of Galactic longitudes in the 0.6-1.0-keV range comparing the model predictions (blue) and the observed profile of the {\it eRosita} bubbles (north and south bubbles shown in orange and red colors, respectively). Panels from left to right correspond to horizontal cuts at Galactic latitudes of $|b|=60^\circ, 50^\circ$, $40^\circ$, and the latitude-averaged profiles. The overall amplitudes and locations of the X-ray shells are well reproduced by the model, though interpretation is complicated by the asymmetry in the data caused by the North Polar Spur observed in the northern X-ray sky.

\newpage

% Figure 5 -- Broad band spectrum including HAWC upper limits, Fermi, eRosita, and microwave spectra. 

% caption only
%{\bf Broad-band spectrum for the Fermi bubbles.} Multi-wavelength spectrum for the region containing the {\it Fermi} bubbles ranging from microwave emission observed by {\it WMAP} and {\it Planck}, GeV gamma rays by {\it Fermi}, and TeV gamma-ray upper limits by {\it HAWC}. For direct comparisons, we computed the simulated spectra within the same extracted regions as the observed spectra, namely, $|l| < 10^\circ, |b|=20^\circ - 40^\circ$ for the gamma-ray bubbles, and $|l| < 25^\circ, |b|=10^\circ - 35^\circ$ for the microwave haze \cite{Ackermann14}. The black and red curves show the predicted emission of the leptonic jet model, in which the primary CR electrons can simultaneously produce the observed gamma-ray emission due to IC scattering and the microwave haze emission by synchrotron radiation, while satisfying the {\it HAWC} constraint.

\newpage

\begin{table}[ht]
\caption{\bf Input and Derived Jet Parameters}
\begin{center}
%\begin{tabular}{|l|l|l|l|}
\begin{tabular}{llll}
\hline
Parameter & Description & Value & Unit \\
\hline
$\eta$ & Density contrast & 0.05 & - \\
$\eta_{\rm e}$ & Energy density contrast & 0.81 & -\\
$e_{\rm jcr}$ & CR energy density & $6.82\times10^{-11}$ & ${\rm erg\ cm}^{-3}$ \\
$v_{\rm jet}$ & Jet speed & 0.025 & $c$ \\
$R_{\rm jet}$ & Radius of cross-section & 0.5 & kpc\\
$t_{\rm jet}$ & Duration of injection & 0.1 & Myr\\
\hline
$n_{\rm ej}$ & Electron number density & 0.1& ${\rm cm}^{-3}$\\
$\rho_{\rm j}$ & Thermal gas densiy & $1.95\times 10^{-25}$ & ${\rm g\ cm}^{-3}$\\
$e_{\rm j}$ & Thermal energy density & $1.29\times 10^{-9}$ & ${\rm erg\ cm}^{-3}$\\
$P_{\rm ke}$ & Kinetic power & $3.08\times 10^{44}$ & ${\rm erg\ s}^{-1}$\\
$P_{\rm th}$ & Thermal power & $7.21\times 10^{42}$ & ${\rm erg\ s}^{-1}$\\
$P_{\rm cr}$ & CR power & $3.82\times 10^{41}$ & ${\rm erg\ s}^{-1}$\\
$P_{\rm B}$ & Magnetic power & $3.31\times 10^{41}$ & ${\rm erg\ s}^{-1}$\\
$P_{\rm jet}$ & Total power & $3.16\times 10^{44}$ & ${\rm erg\ s}^{-1}$\\
$E_{\rm jet}$\footnotemark[1]  & Total injected energy & $1.00\times 10^{57}$ & erg\\
\hline
\multicolumn{4}{l}{\footnotesize \footnotemark[1] The total injected energy by both bipolar jets is $2E_{\rm jet}$.}
\end{tabular}
\end{center}
\label{tbl:jet params}
\end{table}

%\noindent LaTeX formats citations and references automatically using the bibliography records in your .bib file, which you can edit via the project menu. Use the cite command for an inline citation, e.g.  \cite{Hao:gidmaps:2014}.

%For data citations of datasets uploaded to e.g. \emph{figshare}, please use the \verb|howpublished| option in the bib entry to specify the platform and the link, as in the \verb|Hao:gidmaps:2014| example in the sample bibliography file.

%The corresponding author is responsible for submitting a \href{http://www.nature.com/srep/policies/index.html#competing}{competing interests statement} on behalf of all authors of the paper. This statement must be included in the submitted article file.

%Figures and tables can be referenced in LaTeX using the ref command, e.g. Figure \ref{fig:stream} and Table \ref{tab:example}.

\end{document}